\documentclass[twocolumn]{aastex631}

\def\lapp{\ifmmode\stackrel{<}{_{\sim}}\else$\stackrel{<}{_{\sim}}$\fi}
\def\gapp{\ifmmode\stackrel{>}{_{\sim}}\else$\stackrel{>}{_{\sim}}$\fi}
\usepackage{graphicx}
\usepackage{multirow}
\usepackage{color}
\usepackage{amsmath}
\usepackage{soul}
\usepackage{hyperref}

\newcommand{\fluxcgs}{\ensuremath{\mathrm{erg}\,\mathrm{s}^{-1}\,\mathrm{cm}^{-2}}}

\newcommand{\kms}{\ensuremath{\mathrm{km}\,\mathrm{s}^{-1}}}

\shorttitle{}
\shortauthors{}
\begin{document}

\title{Investigation of the non-thermal X-ray emission from the supernova remnant CTB~37B hosting
the magnetar CXOU~J171405.7$-$381031}

\correspondingauthor{Hongjun An}
\email{hjan@cbnu.ac.kr}
\author[0000-0003-0226-9524]{Chanho Kim}
\author[0000-0002-9103-506X]{Jaegeun Park}
\author[0000-0002-6389-9012]{Hongjun An}
\affiliation{Department of Astronomy and Space Science, Chungbuk National University, Cheongju, 28644, Republic of Korea}
\author[0000-0002-9709-5389]{Kaya Mori}
\affiliation{Columbia Astrophysics Laboratory, 550 West 120th Street, New York, NY 10027, USA}
\author[0000-0002-5365-5444]{Stephen P. Reynolds}
\affiliation{Physics Department, NC State University, Raleigh, NC 27695, USA}
\author[0000-0001-6189-7665]{Samar Safi-Harb}
\affiliation{Department of Physics and Astronomy, University of Manitoba, Winnipeg, MB R3T 2N2, Canada}
\author[0000-0002-2967-790X]{Shuo Zhang}
\affiliation{Department of Physics and Astronomy, Michigan State University, East Lansing, MI 48824, USA}

\begin{abstract}
We present a detailed X-ray investigation of a region (S1) exhibiting non-thermal X-ray
emission within the supernova remnant (SNR) CTB~37B hosting the magnetar CXOU~J171405.7$-$381031.
Previous analyses modeled this emission with a power law (PL),
inferring various values for the photon index ($\Gamma$) and absorbing column density ($N_{\rm H}$).
Based on these, S1 was suggested to be the SNR shell, a background pulsar wind nebula (PWN),
or an interaction region between the SNR and a molecular cloud.
Our analysis of a larger dataset favors
a steepening (broken or curved PL) spectrum over a straight PL,
with the best-fit broken power-law (BPL) parameters of $\Gamma=1.23\pm0.23$ and $2.24\pm0.16$ below and above a break at
$5.57\pm0.52$\,keV, respectively. 
However, a simple PL or {\tt srcut} model cannot be definitively ruled out.
For the BPL model, the inferred $N_{\rm H}=(4.08\pm0.72)\times 10^{22}\rm \ cm^{-2}$
towards S1 is consistent with that of the SNR, suggesting a physical association.
The BPL-inferred spectral break $\Delta \Gamma \approx 1$ and hard
$\Gamma$ can be naturlly explained by a non-thermal
bremsstrahlung (NTB) model.  We present an evolutionary NTB model that reproduces the observed spectrum, which indicates the presence of sub-relativistic electrons within S1.
However, alternate explanations for S1,
an unrelated PWN or the SNR shock with unusually
efficient acceleration, cannot be ruled out.  
We discuss these explanations
and their implications for gamma-ray emission from CTB~37B,
and describe future observations that could settle the origin of S1.
\end{abstract}

\bigskip 
\section{Introduction}
\label{sec:intro}
High-energy cosmic rays nearing PeV energies have been suggested to originate from Galactic
sources such as supernova remnants (SNRs) and pulsar wind nebulae (PWNe).
PWNe are believed to primarily accelerate leptons,
while SNRs are thought to be responsible for hadron acceleration.
Evidence for energetic hadrons in several SNRs comes from their
gamma-ray spectra \citep[e.g.,][]{Ackermann2013SNR}.

Three primary radiation mechanisms involving energetic leptons or hadrons are thought to be responsible
for gamma-ray emission from astrophysical objects. Inverse-Compton (IC) scattering
refers to the process where electrons boost the energy of low-energy photons, such as those
from the cosmic microwave background or interstellar radiation field (ISRF), to TeV energies.
Additionally, non-thermal bremsstrahlung (NTB) radiation emitted by energetic electrons can contribute
to the observed gamma-ray emission \citep[e.g.,][]{Chevalier1999,Slane2015}.
On the other hand, the hadronic process entails the collision of high-energy protons,
accelerated by SNR shocks or through interaction between an SNR and a molecular cloud (MC),
with a dense surrounding medium \citep[e.g.,][]{Bykov2000}.
These collisions give rise to neutral pions, which subsequently decay into MeV--TeV gamma rays.

These concurrent leptonic and hadronic processes can coexist within a given source.
Therefore, definitively identifying ``hadronic'' acceleration requires careful
consideration of the aforementioned radiation mechanisms to rule out a purely
leptonic origin for the gamma-ray emission via IC and/or NTB processes.
This necessitates an approach involving the analysis of the multi-wavelength image
and spectral energy distribution (SED), and ultimately, the application of emission models
to the observational data \citep[e.g.,][]{Reynolds2008}.

\begin{figure}
\centering
\includegraphics[width=3.2 in]{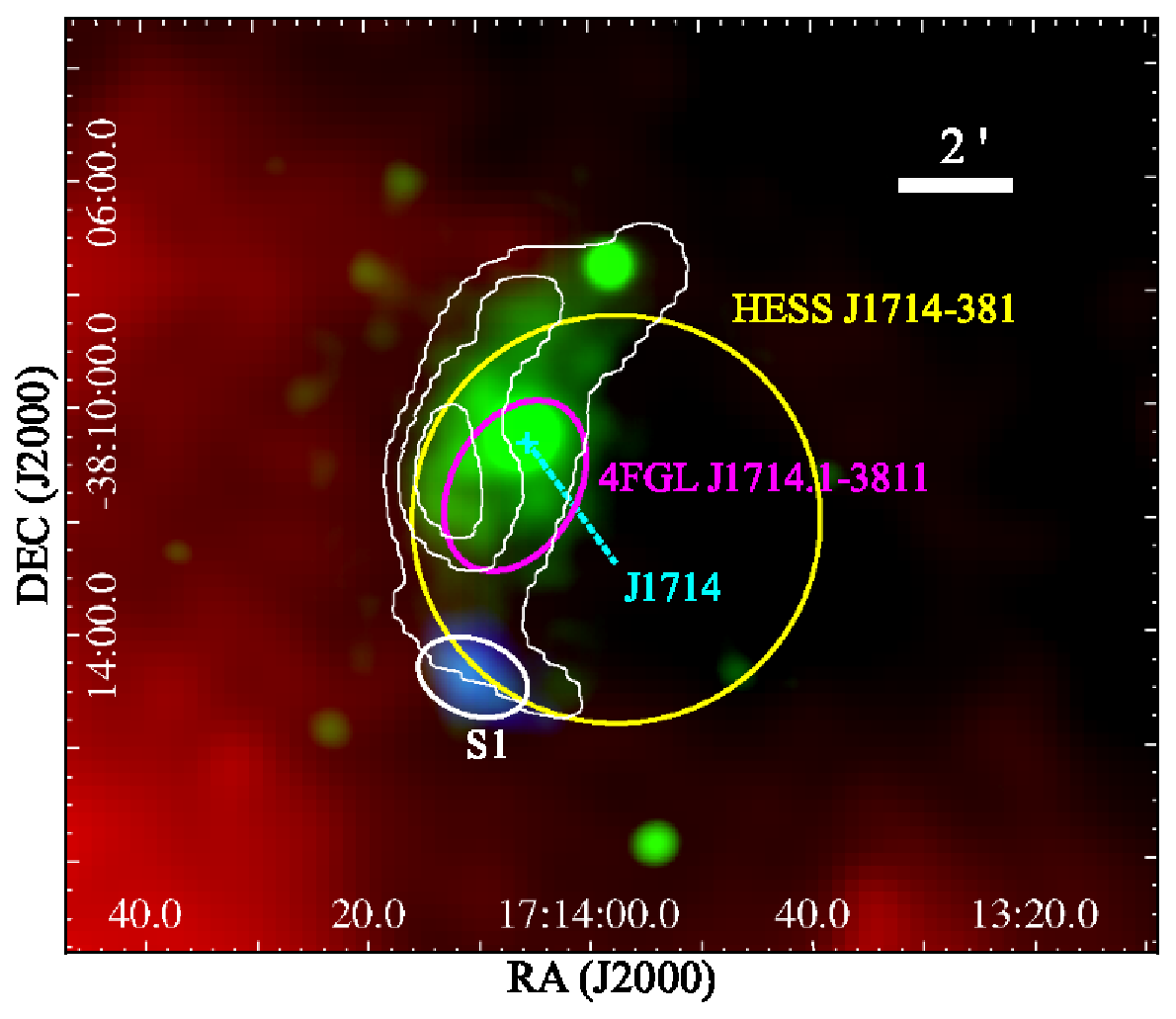} 
\figcaption{A composite image of combined SUMSS (843\,MHz; white contours), Herschel SPIRES (red),
XMM-Newton (1--8\,keV; green) and NuSTAR (3--20\,keV; blue) data of CTB~37B.
The magnetar is marked as J1714 (cyan cross), and our main target ``S1'' is denoted by a white ellipse.
The NuSTAR image was truncated to remove stray-light contamination (e.g., Figure~\ref{fig:fig2} (a)).
The magenta ellipse and the yellow circle show the GeV and TeV counterparts
detected by Fermi-LAT (magenta; 68\% positional uncertainty) and H.E.S.S. (yellow; 1$\sigma$ extension), respectively.
The images were smoothed and logarithmically scaled to improve legibility.
\label{fig:fig1}}
\vspace{-3 mm}
\end{figure}

CTB~37B (G348.7+0.3)\footnote{http://snrcat.physics.umanitoba.ca/SNRtable.php}
is an SNR harboring the bright magnetar CXOU~J171405.7$-$381031 (hereafter J1714)
with a spin period of 3.8\,s, surface dipole magnetic-field strength of $B_s=4.8\times 10^{14}$\,G,
and spin-down power of $\dot E_{\rm SD}=4.2\times 10^{34}\rm \ erg\ s^{-1}$ \citep[][]{Halpern2010,Sato2010}.
The estimated distance to and age of the SNR are 8--13\,kpc
and 650--6200\,yr, respectively \citep[][]{Tian2012,Nakamura2009,Blumer2019}.
The SNR has been well detected across various wavelengths, including radio \citep[][]{Kassim1991},
X-ray \citep[][]{Nakamura2009}, GeV \citep[][]{fermi4fgl}, and TeV bands \citep[][]{Aharonian2008}.
Its radio emission emanates from a shell-like structure east of the magnetar (Figure~\ref{fig:fig1}).
Diffuse X-ray emission was detected surrounding the magnetar; this
X-ray emission region is mostly contained within the radio shell.
The GeV and TeV emissions exhibit significant spatial overlap with both radio and X-ray regions.

Previous TeV observations \citep[e.g., Figure~\ref{fig:fig1}][]{Aharonian2008} of the SNR
favored hadronic processes over leptonic ones.
This is because leptonic scenarios would require an unrealistically low magnetic field strength ($B$)
of $\sim$1$\,\mu$G or an unexpected cutoff in the electron distribution at around 40\,TeV.
\citet{Zeng2017} proposed a lepto-hadronic model where hadronic interactions dominate the TeV emission.
Their model necessitates a high gas density within the shell ($10\rm \ cm^{-3}$) to match
the supernova (SN) energy budget (e.g., $E_{\rm SN}=10^{51}\rm \ erg\ s^{-1}$). This value significantly
exceeds the density inferred from models of the SNR shell's overall X-ray emission
\citep[e.g., $0.5\rm \ cm^{-3}$;][]{Aharonian2008,Blumer2019}.
The origin of this discrepancy in gas density estimations remains unclear in their work.

While the X-ray spectrum of the diffuse emission around J1714 was found to be thermal
\citep[][]{Aharonian2008}, \citet{Nakamura2009} observed non-thermal hard power-law (PL)
emission in the south of the magnetar, and
subsequent investigations by Chandra and XMM-Newton resolved this PL emission to originate from
a compact region at $\sim$4$'$ south of J1714
\citep[S1 in Figure~\ref{fig:fig1}; see also][]{Blumer2019,Gotthelf2019}.
These previous studies reported inconsistent spectral properties for S1,
including photon index ($\Gamma$) and absorbing column density ($N_{\rm H}$) (Section~\ref{sec:sec2_4}).
As a result, the origin of S1's emission remains uncertain. 
While \citet{Nakamura2009} attributed it to the SNR shell, \citet{Gotthelf2019} proposed a background source.
\citet{Blumer2019} considered both a SNR-MC interaction and an unrelated PWN
as possible explanations. The potential association of this non-thermal emission with the SNR holds
significant implications for the radiation processes at play within CTB~37B.

This non-thermal X-ray region S1 could contribute to the total TeV flux through processes
like hadronic interactions or NTB \citep[e.g.,][]{Reynolds2008, Slane2015}.
Based on hard X-ray spectra with $\Gamma<2$ within the SNRs IC~443 and W49B,
\citet{Zhang2018} and \citet{Tanaka2018} suggested NTB emission from them.
Moreover, \citet{Tanaka2018} explored the possibility of gamma-ray emission
through the NTB process in W49B.
However, previous studies on the broadband SED of CTB~37B
have not considered this possibility.

In this paper, we investigate the S1 emission using both archival and
newly acquired X-ray data from XMM-Newton and NuSTAR.
Our analysis methods and results are presented
in Section~\ref{sec:sec2}. We interpret the X-ray data under the framework of
the NTB scenario in Section~\ref{sec:sec3}.
In Section~\ref{sec:sec4}, we discuss the implications of our analysis results.

\newcommand{\marka}{\tablenotemark{\tiny{\rm a}}}
\newcommand{\markb}{\tablenotemark{\tiny{\rm b}}}
\newcommand{\markc}{\tablenotemark{\tiny{\rm c}}}
\begin{table}[t]
\vspace{-0.0in}
\begin{center}
\caption{X-ray data used in this study}
\label{ta:ta1}
\vspace{-0.05in}
\scriptsize{
\begin{tabular}{lcccc} \hline
Observatory & Date (MJD)  &  Obs. ID    & Exposure (ks) \\ \hline
XMM-Newton  & 55273 & 0606020101  & 99/51\marka \\
XMM-Newton  & 55999 & 0670330101  & 8\markb  \\
XMM-Newton  & 57655 & 0790870201  & 24\markb \\
XMM-Newton  & 57806 & 0790870301  & 17\markb \\
NuSTAR      & 57654 & 30201031002 & 79/78\marka  \\
NuSTAR      & 60234 & 40901004002 & 80/78\marka  \\ \hline
\end{tabular}}
\end{center}
\vspace{-0.5 mm}
\footnotesize{
\marka{MOS/PN and FPMA/FPMB for XMM-Newton and NuSTAR, respectively.}\\
\markb{PN only.}\\
}
\end{table}

\section{X-ray Data Analysis}
\label{sec:sec2}
We focus on the S1 emission identified at $\sim$4$'$ south of J1714.
This source was detected in four XMM-Newton and two NuSTAR observations.
NuSTAR detected S1 with high significance ($\ge 10\sigma$) in the 10--20 keV band, confirming
its spectrally hard X-ray emission.
While a Chandra observation also captured this non-thermal source,
its emission is very faint and situated across the chip gap.
Consequently, we have excluded these Chandra data from our analysis.
The X-ray datasets analyzed in this study are listed in Table~\ref{ta:ta1}.

\subsection{Data Reduction}
\label{sec:sec2_1}
The XMM-Newton observations were processed using the pipeline tasks {\tt emproc}
and {\tt epproc} of the XMMSAS software (version 20230412\_1735).
Particle flare contamination was removed from each observation following the standard procedures.
The source was detected by the MOS detector in only one observation, as
the remaining three employed the small-window mode for MOS.
The source was well-detected in all four PN datasets.
The NuSTAR observations were processed using the {\tt nupipeline} script integrated in HEASOFT\,v6.32.
While the source was well detected in the NuSTAR observations (e.g., Figure~\ref{fig:fig2} (a)), its proximity
to stray-light patterns hindered reliable background estimation from the in-flight data,
necessitating a careful examination of the background during analysis.
The net exposure times after these initial cleaning steps are presented in Table~\ref{ta:ta1}.

\subsection{Image Analysis}
\label{sec:sec2_2}
We used the {\tt eimageget} script of SAS for each XMM-Newton observation to create
a background-subtracted and vignetting-corrected image.\footnote{https://www.cosmos.esa.int/web/xmm-newton/sas-thread-images}
These images were combined to produce a 1--8 keV image.
3--20\,keV NuSTAR images were generated
with background subtracted using {\tt nuskybgd} simulations \citep[][]{Wik2014} and
exposure corrections applied. We then combined these NuSTAR images
with the XMM-Newton image alongside an IR image measured by Herschel SPIRES.
These energy bands were selected to optimize signal-to-noise ratio.
We additionally displayed radio contours obtained from the SUMSS data \citep[][]{Mauch2003}.
The resultant radio-to-X-ray composite image is shown in Figure~\ref{fig:fig1}.
Notably, the radio contours overlap well with the XMM-Newton image and
the IR emission seems to delineate the radio SNR shell to the east.

The X-ray image, encompassing J1714, SNR emission surrounding it, and the southern
non-thermal emission (`S1'), appears to exhibit an overall morphology similar to the radio shell.
In both the XMM-Newton and NuSTAR data, S1 manifests as extended emission with $R\sim1'$.
The radio and X-ray emissions significantly overlap with the GeV and TeV emissions measured
by Fermi-LAT \citep[4FGL\,J1714.1$-$3811;][]{fermi4fgl}
and H.E.S.S \citep[HESS\,J1713$-$381;][]{HESS2018}.
It is worth noting that the GeV emission remains unresolved and seems to originate from within the SNR shell,
whereas the TeV emission exhibits extension with a Gaussian width ($\sigma$) of 5.5$'$, covering
a large region containing both the shell and S1.

\subsection{Timing Analysis for J1714}
\label{sec:sec2_3}
\begin{figure}
\centering
\hspace{-2.3 mm}
\includegraphics[width=2.84 in]{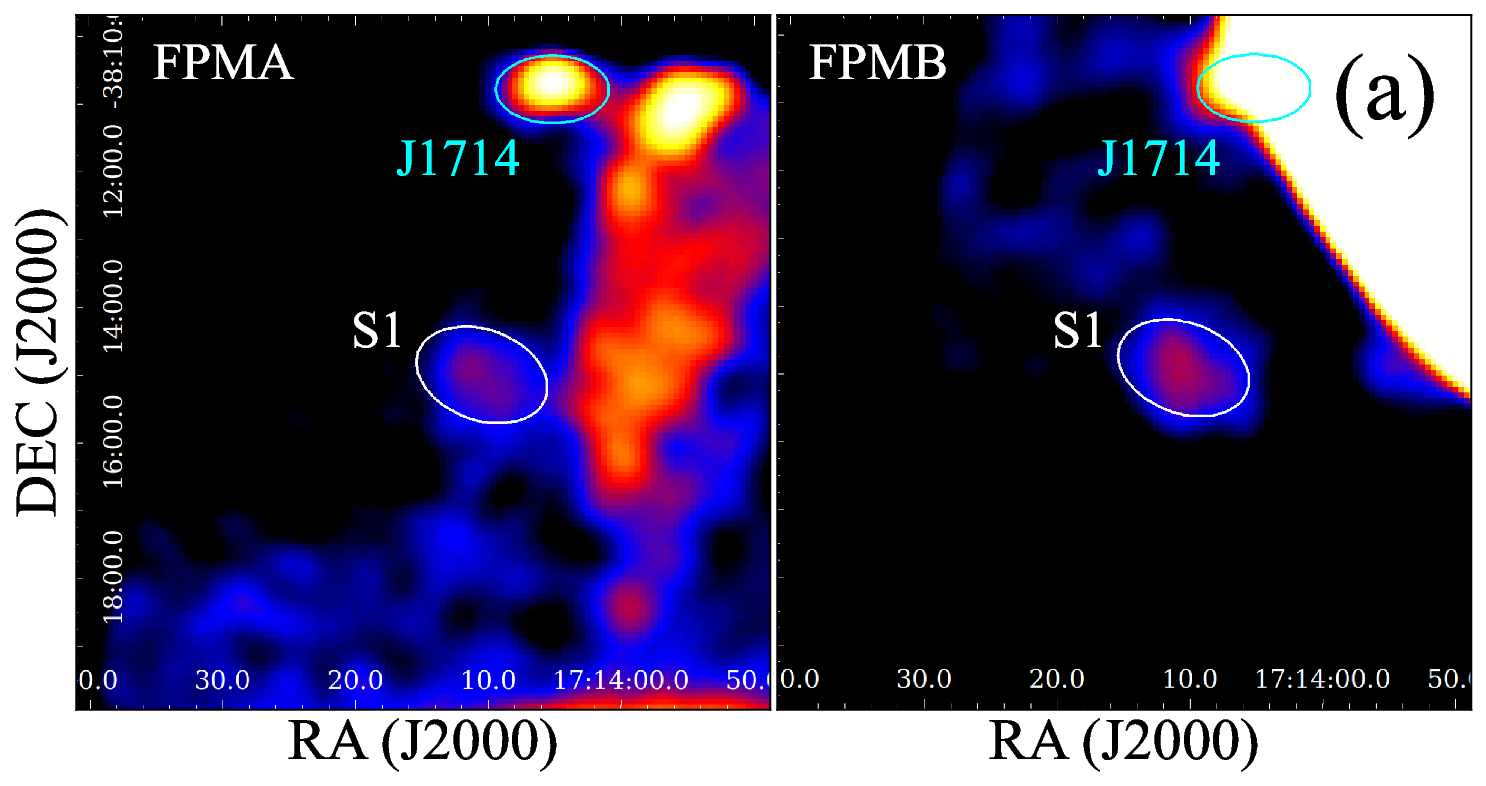} 
\includegraphics[width=3.2 in]{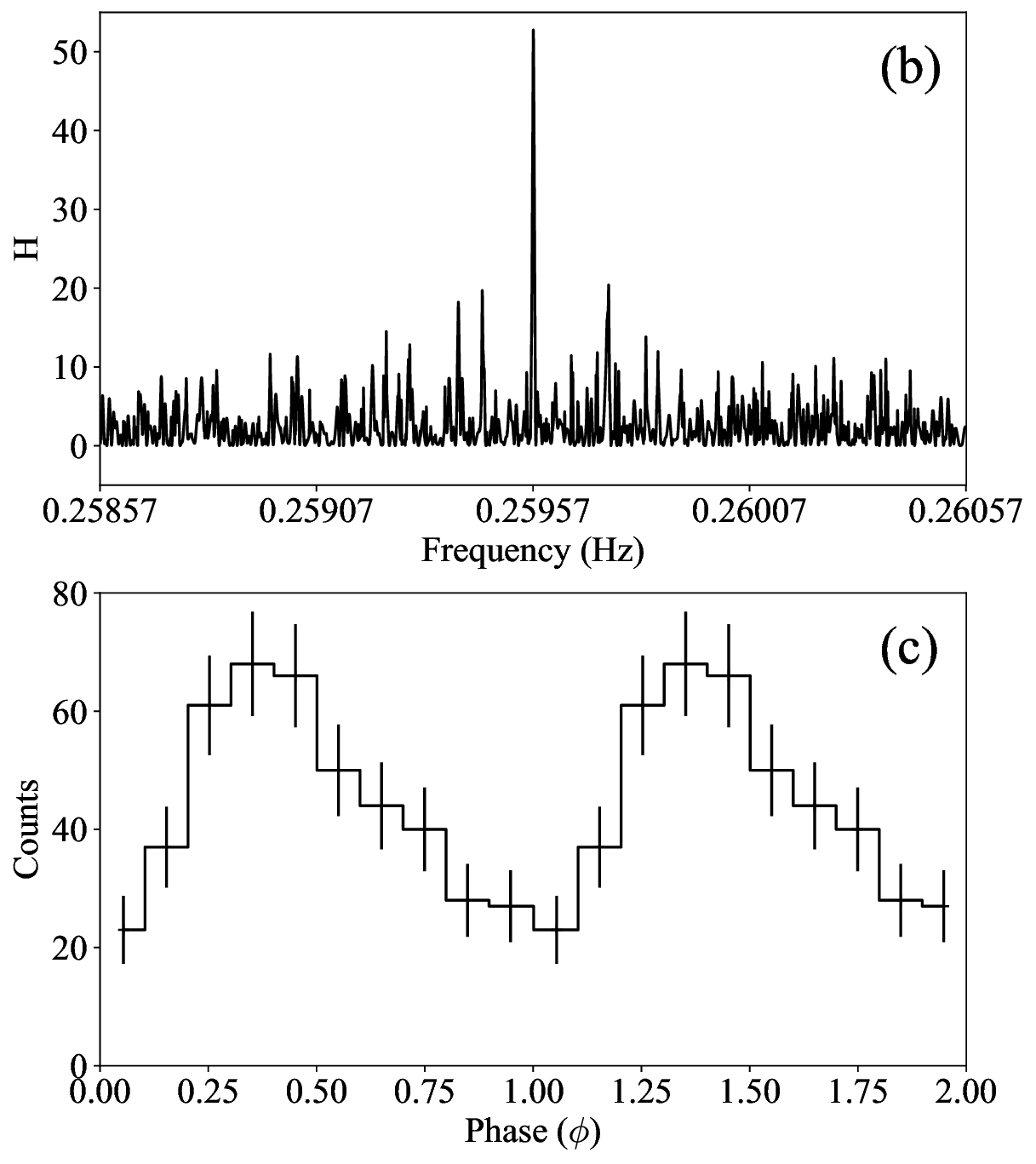} 
\vspace{-5 mm}
\figcaption{(a) 3--20\,keV NuSTAR FPMA (left) and FPMB (right) images made with the 2023 observation.
While J1714 appears to be heavily contaminated by stray light in the FPMB image,
it was detected outside the contamination in FPMA.
We smoothed and logarithmically scaled the images to enhance legibility.
(b and c) Result of our pulsation search (b) and background-subtracted 1.6--5\,keV pulse profile (c)
of J1714 measured using the 2023 NuSTAR data.
\label{fig:fig2}}
\vspace{-2 mm}
\end{figure}

Although the magnetar J1714 is not our main target and its emission was heavily contaminated
by stray light (bright regions except for J1714 and S1 in Figure~\ref{fig:fig2} (a)) in the NuSTAR observations,
we carry out a timing analysis with the new data to ascertain if there have been any significant
changes in the magnetar's rotation since the last measurement \citep[][]{Halpern2010,Sato2010,Gotthelf2019}.
During the new observation, the magnetar was observed at $\sim5'$ off-axis position near the edge of the detector,
resulting in a distorted event distribution and reduced counts (Figure~\ref{fig:fig2} (a)).
Moreover, due to the bright stray-light pattern overlapping with J1714 in the FPMB data,
we relied solely on the FPMA data for the timing analysis.
We selected source events within a $50''\times30''$ (radii) elliptical region, barycenter-corrected their arrival times
using (RA, Decl)=($258.5239057^\circ$, $-38.1752758^\circ$) (J2000),
and conducted an $H$-test \citep[][]{drs89} around the expected period based on previous measurements.
The spin period derivative ($\dot P$) was held fixed at a previously reported value of $5\times 10^{-11}\rm \ s\ s^{-1}$.

Our analysis successfully detected pulsations at low energies (1.6--5\,keV),
revealing a period of $P$=3.852450(5)\,s on MJD~60233 (Figure~\ref{fig:fig2} (b) and (c)).
Unfortunately, owing to limited statistics resulting from
a large off-axis angle and elevated background due to stray light,
we were unable to confirm the previously reported phase
reversal of the pulse profile at higher energies \citep[$>$6\,keV;][]{Gotthelf2019}
with the new data.
However, the obtained $P$ value aligns with the trend presented in \citet{Gotthelf2019};
by comparing our result with the previous Chandra measurement on MJD~54856,
we estimated an average $\dot P$ of $6\times 10^{-11}\rm \ s\ s^{-1}$, falling within the previously reported range
of (5--7)$\times 10^{-11}\rm \ s\ s^{-1}$.

\subsection{Spectral Analysis of the Emission from S1}
\label{sec:sec2_4}

\begin{figure*}
\centering
\begin{tabular}{ccc}
\includegraphics[width=2.25 in]{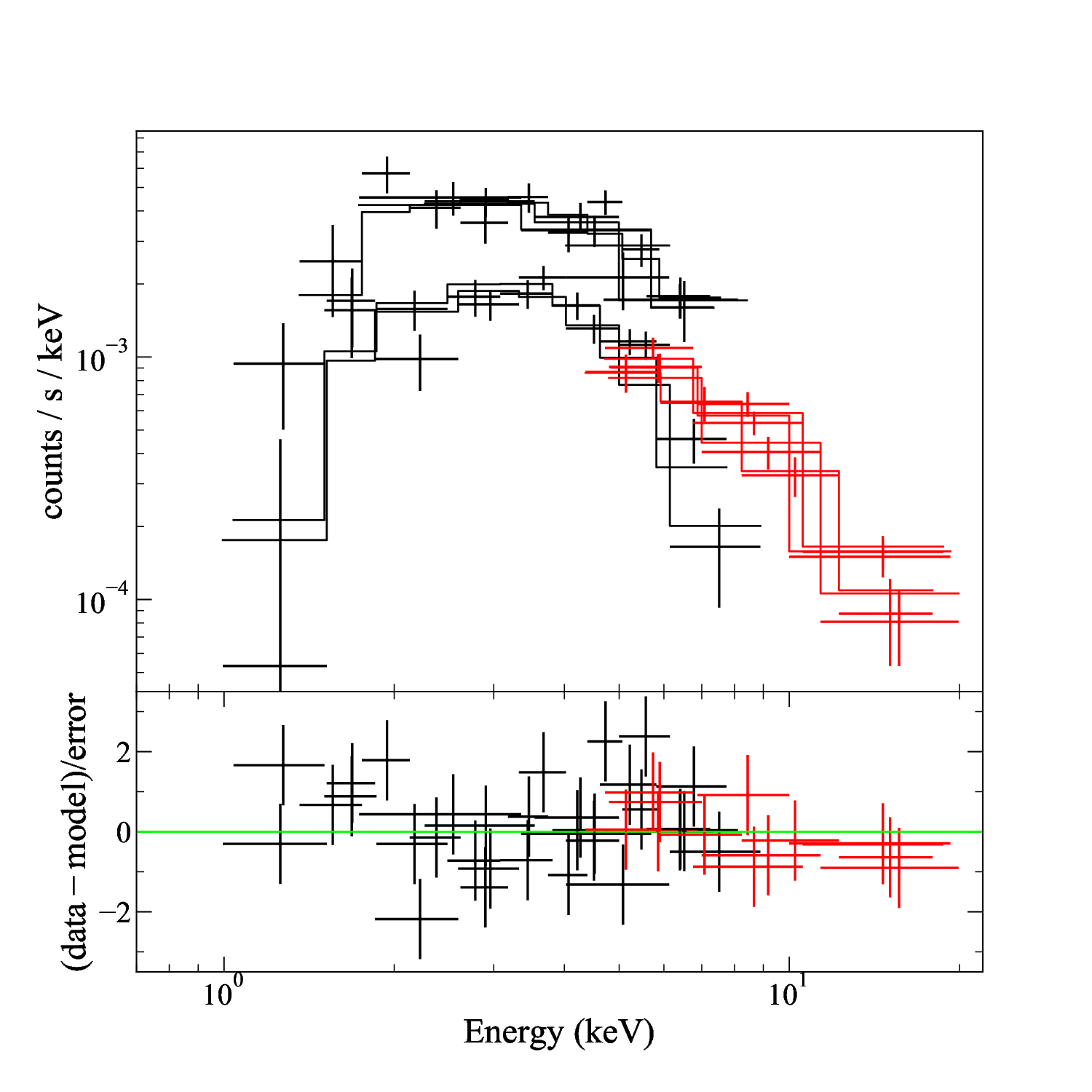} &
\includegraphics[width=2.25 in]{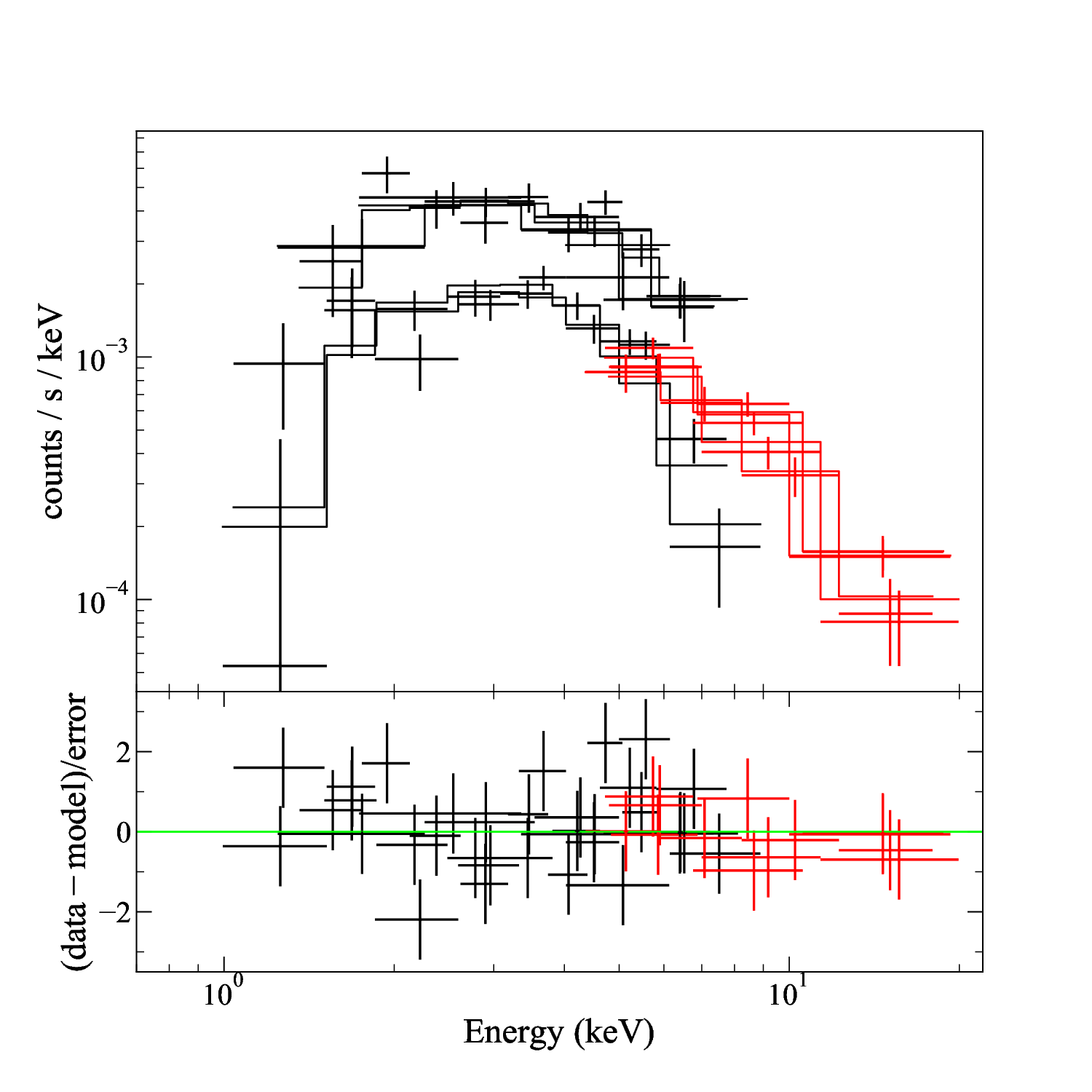} &
\includegraphics[width=2.25 in]{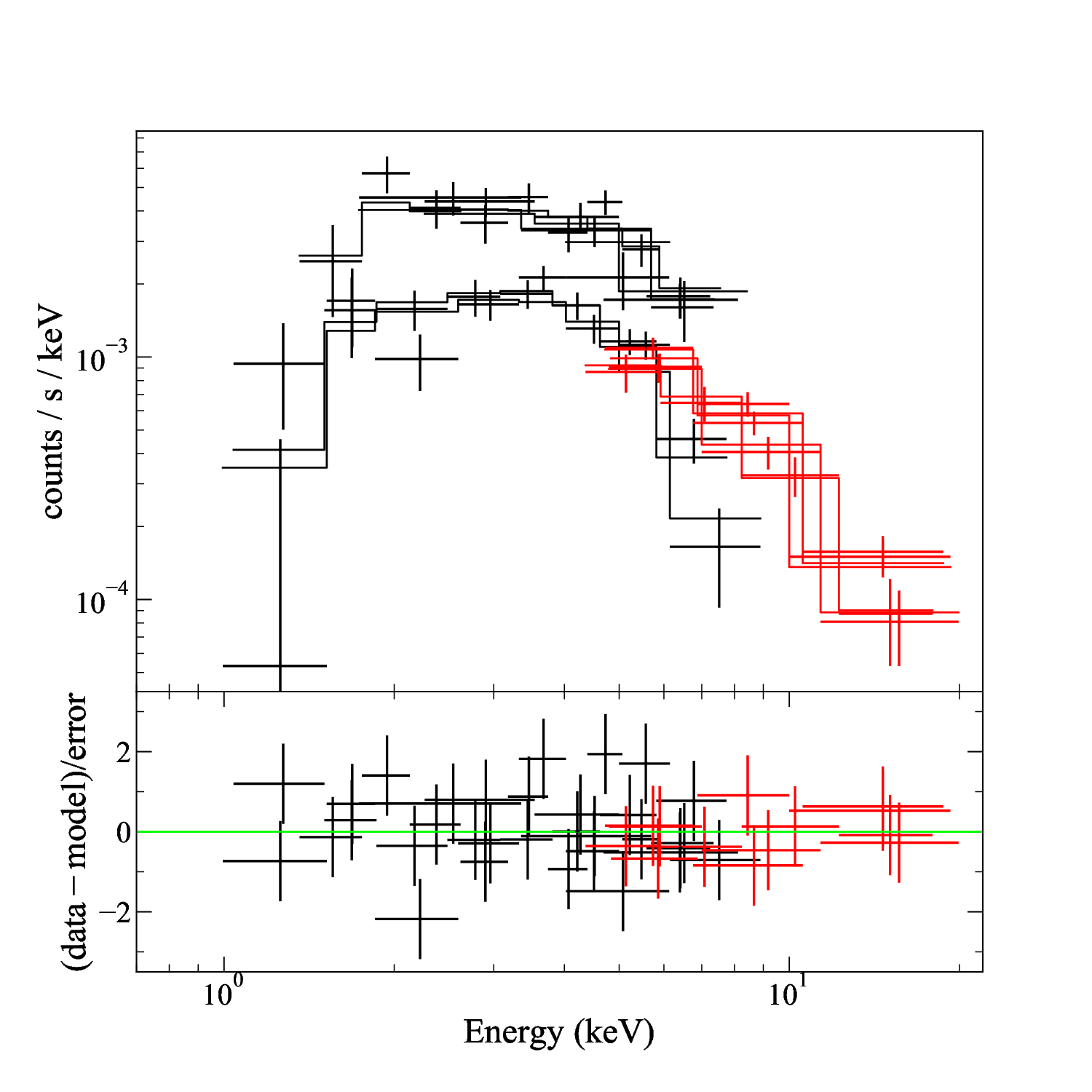} \\
\end{tabular}
\vspace{0mm}
\figcaption{1--20\,keV X-ray spectra of the S1 region measured
by XMM-Newton (black) and NuSTAR (red), and the best-fit PL (a), {\tt srcut} (b), and BPL models (c).
The bottom panels display residuals after subtracting the best-fit model from the data.
\label{fig:fig3}}
\end{figure*}

Several previous studies have extensively characterized the X-ray spectrum of S1 \citep[][]{Nakamura2009,Blumer2019,Gotthelf2019}.
\citet{Nakamura2009} analyzed a large region encompassing S1 in Suzaku XIS data,
wherein they employed a model to fit the spectrum while accounting for contamination
from J1714 and the thermal shell of the SNR. 
Regarding the S1 emission, they derived $\Gamma=1.5\pm0.4$ for $N_{\rm H}=3.5^{+0.5}_{-0.7}\times 10^{22}\rm \ cm^{-2}$
using the {\tt angr} abundances \citep[][]{angr89}.
While this approach provided insights into the broader region,
it was susceptible to contamination from other sources.
XMM-Newton's high angular resolution facilitated a more precise measurement of S1's spectrum,
minimizing contamination from other sources. \citet{Blumer2019} analyzed XMM-Newton MOS data (Obs. ID 0670330101)
and reported $\Gamma=1.3\pm0.3$ and $N_{\rm H}=3.1^{+0.9}_{-0.8}\times 10^{22}\ \rm cm^{-2}$ for S1,
while \citet{Gotthelf2019} analyzed the same MOS data jointly with
NuSTAR spectra (Obs. ID 30201031002), finding a steeper $\Gamma$ of $2.2^{+0.6}_{-0.5}$
and higher absorption with $N_{\rm H}=(11\pm4)\times 10^{22}\ \rm cm^{-2}$.
It should be noted that these $N_{\rm H}$ values were determined employing the {\tt wilms}
abundance model \citep[][]{wam00}.

\begin{table*}[t]
\vspace{-0.0in}
\begin{center}
\caption{Results of joint fits of the XMM-Newton and NuSTAR spectra}
\label{ta:ta2}
\vspace{-0.05in}
\scriptsize{
\begin{tabular}{lccccccc} \hline
Model &  Energy range & $N_{\rm H}$ & $\Gamma/\alpha$\marka & $E_{\rm brk}$ & $\Gamma_2$\markb & Flux\markc & $\chi^2$/dof \\ \hline
 & (keV)  & ($10^{22}\rm cm^{-2}$)  & & (keV) &  &  & \\ \hline
PL  & 0.3--20\,keV & $6.52\pm0.52\pm0.11$ & $1.95\pm0.09\pm0.03$ & $\cdots$ & $\cdots$ & $3.18\pm0.20\pm0.01$ & 524/549 \\
{\tt srcut}  & 0.3--20\,keV & $5.91\pm0.45\pm 0.04$ & $0.30$ & $4.31\pm1.70\pm 0.05$ & $\cdots$ & $35.00\pm 8.11\pm 1.33$ & 520/549 \\
BPL & 0.3--20\,keV & $4.08\pm0.72\pm0.07$ & $1.23\pm0.23\pm0.02$ & $5.57\pm0.52\pm0.06$ & $2.24\pm0.16\pm0.06$ & $2.79\pm0.21\pm0.02$ & 510/547 \\ \hline
\end{tabular}}
\end{center}
\vspace{-0.5 mm}
\footnotesize{Note. The statistical and systematic uncertainties, reported as
the first and second errors, respectively, are at the 1$\sigma$ confidence level.\\
\marka{X-ray photon index ($\Gamma$) for the PL and BPL models, and radio spectral index ($\alpha$) for the {\tt srcut} model. We held $\alpha$ fixed at 0.3.}\\
\markb{Photon index above the break energy $E_{\rm brk}$.}\\
\markc{Absorption-corrected 2--10\,keV flux in units of $10^{-13}$\,\fluxcgs\ for the PL and BPL models, and flux density at 1\,GHz in units of $\mu$Jy for the {\tt srcut} model.}\\
}
\end{table*}

Given the ongoing debate surrounding the origin and spectral characteristic of the S1 emission,
we acquired a new NuSTAR observation and reanalyzed the existing XMM-Newton and NuSTAR data
(Table~\ref{ta:ta1}).
While the previous XMM-Newton studies utilized only the MOS data,
the source was well detected by XMM-Newton PN. Expanding the dataset to include the PN data can
potentially improve the previous characterization of the S1 spectrum.
The source spectra were extracted using an elliptical region of
$60''\times40''$ (radii) centered at (RA, Decl)=($258.5423^\circ$, $-38.2471^\circ$) from both XMM-Newton and NuSTAR data,
as depicted in Figures~\ref{fig:fig1} and \ref{fig:fig2} (a).
While the in-flight data around the source region can effectively
represent the background in the XMM-Newton data,
the complex stray-light pattern in the NuSTAR data (e.g., Figure~\ref{fig:fig2} (a))
poses challenges for background estimation.
To address this issue, we conducted {\tt nuskybgd} simulations to estimate the NuSTAR
background, utilizing source-free regions while excluding the stray-light patterns.
These simulations provided estimates of background contributions to the source-region spectra.
For XMM-Newton data, we extracted background from 45$''$ radius circles
located 150$''$ west of the source, for both MOS and PN data.
These background regions were chosen to be on the same detector chips as the source, avoiding S1 and the SNR shell.
Alternative background regions (south or west of S1) were tested,
and we found that the results do not alter significantly (see below).

We initially performed independent fits to the XMM-Newton and NuSTAR spectra.
We collected 6700/3400 and 3300/1600 counts within the source/background regions
from the XMM-Newton (0.3--10\,keV; all observations combined) and NuSTAR data (3--20\,keV; all observations combined).
As reported by \citet{Blumer2019}, the XMM-Newton data favor a hard PL model
with $\Gamma=1.35\pm0.17$ and $N_{\rm H}=(4.38\pm0.65)\times 10^{22}\rm \ cm^{-2}$.
Conversely, the NuSTAR data are well-fit by a softer PL model with $\Gamma=2.06 \pm 0.09$ for a fixed
$N_{\rm H}$ of $4.38\times 10^{22}\rm \ cm^{-2}$.
Optimizing $N_{\rm H}$ for the NuSTAR fit results in a softer $\Gamma$ of $2.31\pm0.18$ with
a higher $N_{\rm H}$ value of $(8.46\pm 2.88)\times 10^{22}\rm \ cm^{-2}$.
For Galactic absorption, we employed the {\tt wilms} abundances
and {\tt vern} cross section \citep[][]{vfky96}.
We checked systematic effects on the $\Gamma$ measurements due to background selection
by employing five different background regions for each dataset.
Depending on the background selection, the $\Gamma$ values inferred from the XMM-Newton
and NuSTAR data varied by $\sim 0.04$ and $\sim 0.05$, respectively.
These are smaller than the statistical uncertainties.

The difference in the $\Gamma$ values obtained from the XMM-Newton and NuSTAR analyses
suggests a potential spectral break or curvature. Consequently,
we jointly fit the combined XMM-Newton and NuSTAR spectra using both an absorbed PL and
a broken PL (BPL) model (Figure~\ref{fig:fig3}). To consider the possibility of synchrotron emission from the SNR shock in the cut-off regime, we also employed the {\tt srcut} model \citep[][]{Reynolds1999}.
This analysis included two MOS spectra (MOS1 and MOS2), four PN spectra,
and four NuSTAR spectra (FPMA and FPMB). A cross-normalization factor was applied to
each dataset, with the value fixed to 1 for the MOS1 spectrum.
The best-fit values for these factors were found to be consistent with 1 within uncertainties.
Detailed values and uncertainties for the spectral parameters are presented in Table~\ref{ta:ta2}.
Systematic uncertainties due to background selection were estimated and reported in the table.

The PL model prefers a large $\Gamma$ of $1.95\pm0.09$ with a high $N_{\rm H}=6.52\pm0.52\rm \ cm^{-2}$.
These parameters are statistically consistent with those reported by \citet{Gotthelf2019}.
The $N_{\rm H}$ value differs at a $>3\sigma$ level from those measured towards the SNR shell
\citep[e.g., $4.3^{+0.2}_{-0.1}\times10^{22}\rm \ cm^{-2}$;][]{Blumer2019}
and J1714 \citep[(3.6--4.0)$\times 10^{22}\rm \ cm^{-2}$;][]{Gotthelf2019}.
Despite an adequate fit to the data ($\chi^2$/dof=524/549), the residuals of the PL model exhibit
a small downward slope with increasing energy and an excess at 5--6\,keV where the BPL model predicts a break.
The {\tt srcut} model, with the previously reported $\alpha$ value of 0.3
\citep[although this is for the entire SNR;][]{Kassim1991},
also yields an acceptable fit to the data without overpredicting
the measured 1\,GHz flux density of the entire SNR.
Similar to the PL model, this model requires a high $N_{\rm H}$. Changing $\alpha$ to a larger value, e.g., 0.5 as observed in other radio SNRs, results in an $E_{\rm brk}$ of $\sim$10\,keV, a flux density of $\sim$mJy, and a $N_{\rm H}$ of $6.2\times 10^{22}\rm \ cm^{-2}$.

$F$-tests indicate that the BPL model provides statistically significant
improvements over the PL and the {\tt srcut} models with $F$-test probabilities of $5\times 10^{-4}$ and $5\times 10^{-3}$, respectively.
This finding remains valid even after excluding a few PN observations wherein
a significant portion of S1 fell on bad pixels.
The BPL fit indicates that the $N_{\rm H}$ value towards S1 is consistent with
those for the SNR shell and J1714,
and the spectrum exhibits a break at 6\,keV with $\Delta \Gamma\approx1$.

\section{NTB model for the non-thermal X-ray emission from S1}
\label{sec:sec3}
Building on our description of S1's X-ray emission with a BPL model,
we investigate the origin of this non-thermal component. The association of S1 with
the SNR shell may suggest particle acceleration resulting from the interaction
of an MC with the SNR shock \citep[][]{Blumer2019}.
While not explicitly identified in previous work,
MC\,3251, located at an estimated distance of 8.1--8.6\,kpc
\citep[][]{Miville2017}, is a plausible candidate for the MC.
Theoretical studies have demonstrated that such shock propagation in an MC can accelerate thermal electrons in weakly
ionized regions to non-thermal energies \citep{Bykov2000}. 
The observed hard photon index and strong $\Delta \Gamma \approx 1$ spectral break
(Section~\ref{sec:sec2_4}) favor the NTB process as the dominant X-ray emission mechanism for S1.
In this section, we construct an evolutionary NTB model and apply it to the X-ray emission from S1.

\begin{figure*}[t]
\centering
\begin{tabular}{cc}
\includegraphics[width=3.27 in]{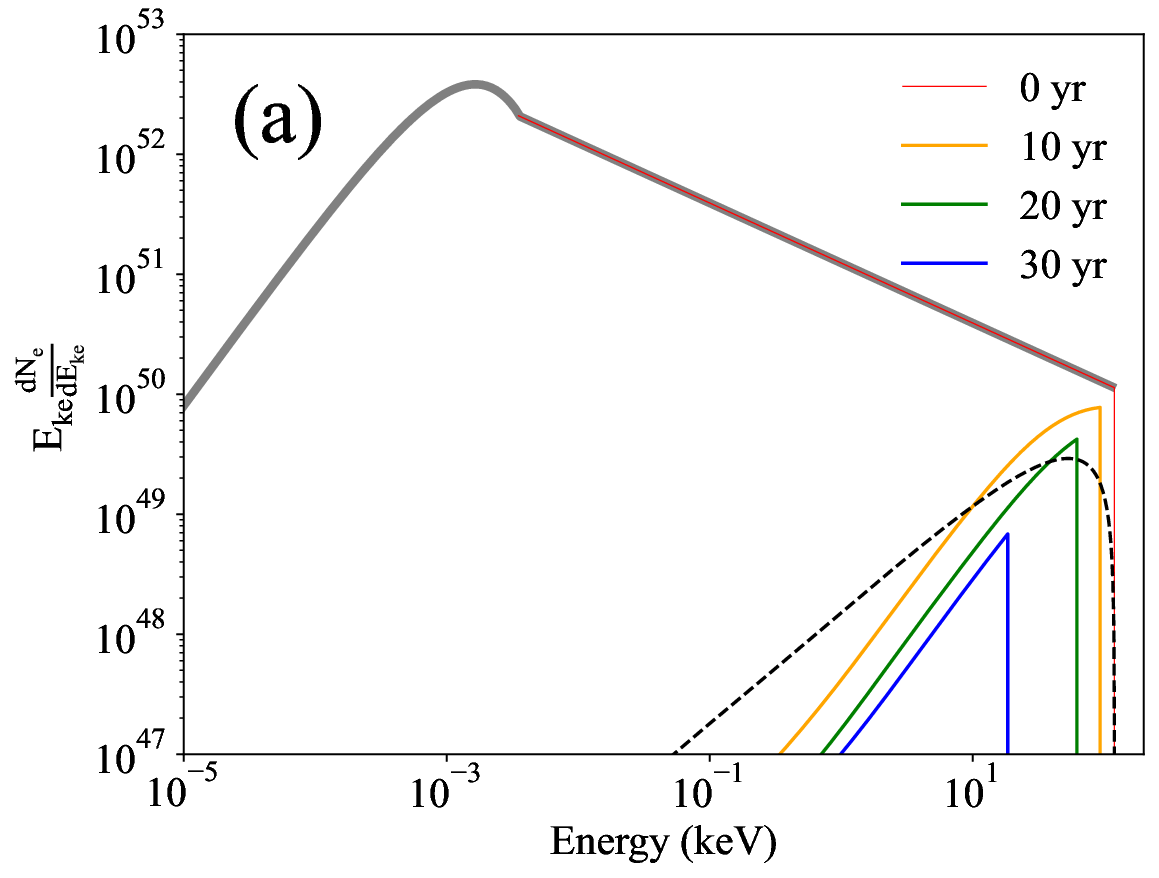} &
\includegraphics[width=3.4 in]{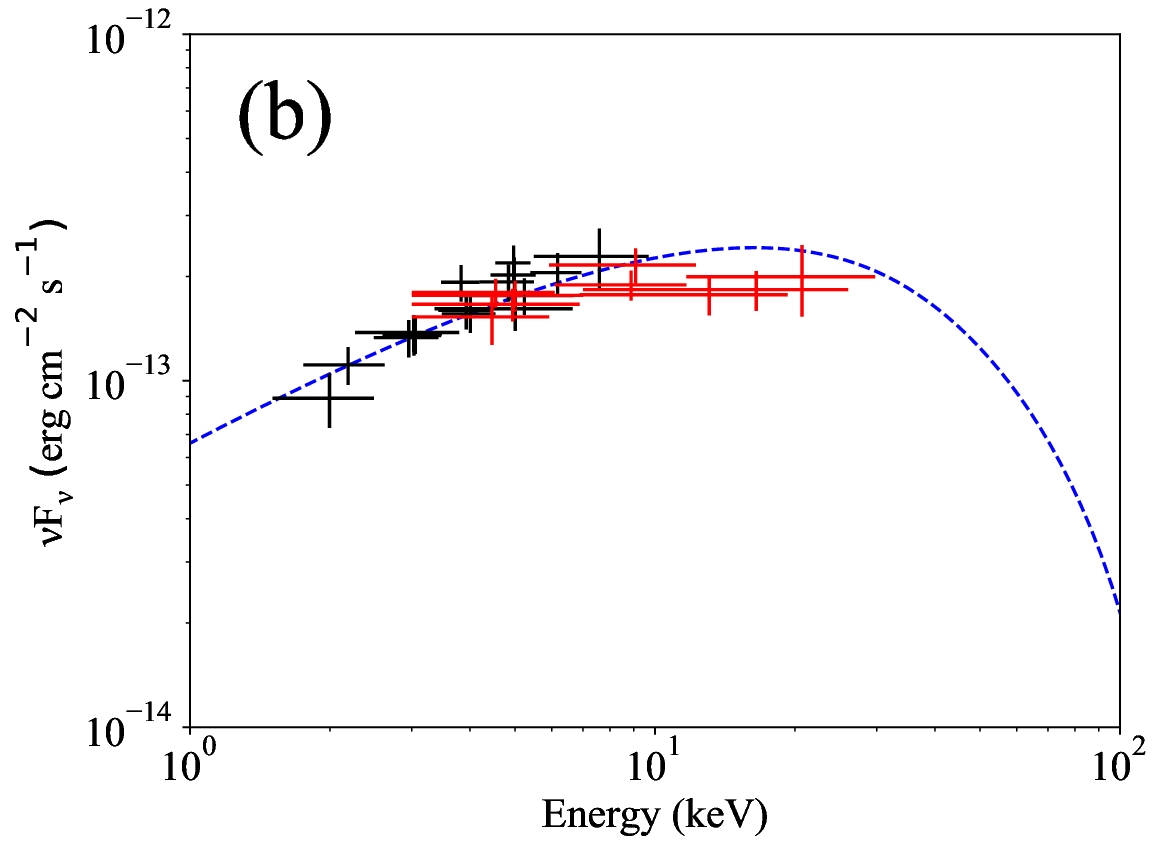} \\
\end{tabular}
\figcaption{(a) Evolution of an electron distribution over 55\,yr for $s=1.5$, and initial $n_e=39\rm \ cm^{-3}$ and $n_{e,b}=81\rm \ cm^{-3}$ (Table~\ref{ta:ta3}). $n_e$ decreases to $\sim 0$ over $\tau_{\rm age}$ while $n_{e,b}$ increase to $120\rm \ cm^{-3}$ as the non-thermal electrons cool (see text).
The gray solid curve displays the summed distribution of the background ($n_{e,b}$; Maxwellian with $kT=1$\,eV) and injected ($n_e$; PL) electrons.
The other solid curves display the time evolution of the PL distribution (young to old from red to purple),
and the black dashed line shows the sum of the distributions (scaled with $dt$).
(b): X-ray emission SED (data points) of S1 and our NTB model computation (dashed line).
\label{fig:fig4}}
\vspace{0mm}
\end{figure*}

\subsection{NTB Emission from Energetic Electrons}
\label{sec:sec3_1}
NTB emission from electrons with energies higher than those in the background plasma, has been
proposed as a source of hard X-ray emission in SNRs by several
authors \citep[e.g.,][]{Tanaka2018, Zhang2018}. Here we summarize some of the properties of
this process. We consider a background plasma composed of ions (density $n_i$)
and electrons (density $n_{e,b}$) in thermal equilibrium at temperature
$T$. To this background, we add a suprathermal electron population with kinetic
energies $E_{ke} \gg E_{\rm th} \sim kT$ and density $n_e$, satisfying
\begin{equation}
n_{e,b} + n_e = 1.2 n_i.
\label{conserv}
\end{equation}
The suprathermal electrons interact with background electrons and ions
on different timescales \citep[e.g.,][]{Spitzer1978},
and radiate bremsstrahlung photons. Electrons will share energy among
themselves on a timescale
\begin{equation}
t_{ee} \cong \frac{1.24 \times 10^{-18}}{\lambda_{ee} n_{e,b}}
\left( \frac{2 E_{ke}}{m_e} \right)^{3/2},
\label{tee}
\end{equation}
where $\lambda_{ee}$ is the Coulomb logarithm, typically $\sim
30$ for $E_{ke} = 10$\,keV, $kT=1$\,eV, and $n_{e,b} = 80\rm \ cm^{-3}$ (as we
find below).
This timescale also approximates the characteristic cooling time for high-energy
electrons due to energy transfer to the thermal electron pool (i.e., $t_{ee} \cong E_{ke}/{\dot{E}_{ke}}$).

In the non-relativistic regime,
the electron energy evolution can be determined analytically \citep{Vink2008}:
\begin{equation}
\label{Ecoul}
E_{ke}(t)^{1.5}=E_{ke}(0)^{1.5}-1.16\times 10^{-5}\lambda_{ee} n_{e,b} t.
\end{equation}
This describes the evolving energy of an electron with initial energy $E_{ke}(0) \gg\ kT$
interacting with a much larger pool of background electrons.
This energy transfer is commonly referred to as Coulomb losses,
although energy remains in the fluid.  In the absence of any additional
acceleration processes such as turbulent acceleration, 
the electron energy distribution evolves over time as lower-energy (suprathermal) electrons cool
and successively disappear into the background plasma.
Consequently, for the electron distribution of
age $t$, there is an energy $E_C$ ($E_{ke}$ with $t = t_{ee}$ in Eq.~(\ref{Ecoul}))
below which the suprathermal electrons are steeply depleted.
This depletion produces a break in the distribution at $E_C$ which rises with time.
Figure~\ref{fig:fig4} (a) illustrates this effect,
showing the evolution of the electron distribution for a PL index $s=1.5$
and $n_{e,b}=81\rm \ cm^{-3}$.

Electrons also scatter off ions with a typical timescale $t_{pe} =
(n_{e,b}/n_i) t_{ee}$ \citep[][]{Spitzer1978}. While these scatterings produce
bremsstrahlung photons, the radiative energy losses are significantly smaller
than those due to Coulomb interactions with background electrons \citep[][]{Petrosian2001}.
An individual electron with $E_{ke}$ emits a bremsstrahlung
photon spectrum that is independent of energy up to $h\nu_{\rm max} \sim E_{ke}$.
Photons of energy $E_\gamma \equiv h\nu$ are typically produced by electrons
with energies several times larger. Consequently, the photon spectrum $N(E_\gamma)$ closely
resembles the electron distribution.

\subsection{NTB Modeling of the X-ray Spectrum of S1}
\label{sec:sec3_2}
We consider a simplified scenario in which the SN that produced
the shell SNR CTB~37B and J1714 sent a blast wave into a
small region of higher density (S1).
This shock accelerated electrons to $\sim$100 keV,
adequate to produce bremsstrahlung photons up to $\sim$20\,keV.
The shock injected a total amount of energy $W_{e,S1}$ to electrons,
proportional to the transverse extent of S1 ($f_{\Omega, S1}$; solid angle fraction of S1):
\begin{equation}
W_{e,S1} = \eta f_{\Omega, S1} E_{\rm SN},
\label{Einj}
\end{equation}
where $\eta$ is the fraction of shock energy converted to electron acceleration.
We describe the electron distribution with a Maxwellian with $kT \sim 1$\,eV (thermal background)
with an attached non-thermal tail (shock-accelerated population). This postshock non-thermal electron distribution,
\begin{equation}
\frac{dN_e}{dE_{ke}dt}=\frac{n_e V_{S1}(s-1)}{(E_{ke,\rm min}^{1-s} - E_{ke,\rm max}^{1-s})\tau_{\rm age}} {E}_{ke}^{-s}
\label{ntbdnde}
\end{equation}
with $V_{S1}$ and $\tau_{\rm age}$ being the volume and lifetime of S1,
evolves due to Coulomb losses as described by Equation~(\ref{Ecoul}).
We ignore the possibility of turbulent reacceleration behind the shock.

To obtain the total photon spectrum at a given age,
we integrate over the shocked region using a one-dimensional geometry.
Assuming the transverse area of the source $A$ ($60''\times 60''$ corresponding to $2.6\times2.6\rm \ pc^2$ for an assumed distance of 9\,kpc),
we divide the emitting region ($V_{S1}$) into discrete volumes $\Delta V = A dz$
where $dz = v_{sh} dt$ and $v_{sh}$ is the shock velocity; we assume it to be 900\,\kms\ as inferred for the SNR shock \citep[][]{Blumer2019}. We calculate $E_{ke}(t)$, the electron distribution, and the emission spectrum for each
volume (time step $dt$). Summing over these spectra yields the total spectrum for the source age 
$\tau_{\rm age}$.

We use energetic considerations to proceed.  First, we assume the total
energy in accelerated particles (non-thermal electrons and ions) is
small enough that the test-particle result for the particle
distribution for diffusive shock acceleration (DSA) applies. This yields $s = 3/2$ for an initial
shocked-particle distribution $N({\bf p}) \propto {\bf p}^{-4}$.
For $s < 2$, most energy is at the higher end. Therefore,
the total energy constraint primarily affects $E_{ke,{\rm max}}(0)$.
The $E_{ke,\rm max}(0)$ value determines the cutoff energy in the electron distribution (Eq.~(\ref{Ecoul})), which corresponds to the energy above which the photon spectrum steepens. Consequently, high values of $E_{ke,\rm max}(0)$ overpredict the observed X-ray spectrum above the spectral break at $\sim$6\,keV.
To reproduce the observed X-ray break,
the initially most energetic electrons (having $E_{ke,\rm max}(0)=120$\,keV; Table~\ref{ta:ta3})
should cool to $\sim$30\,keV. 
Equation~(\ref{Ecoul}) then gives $n_{e,b}\tau_{\rm age}\sim 3300\rm\ cm^{-3}\ yr$.
A low $n_e$, limited by the SN energy budget, necessitates a high $n_i$ to
explain the observed flux since the NTB flux scales as $\propto n_i n_e V_{S1}$.
This leads to a high $n_{e,b}$ (Eq.~(\ref{conserv})) and a correspondingly short $\tau_{\rm age}$.

We optimized model parameters to reproduce the X-ray spectrum, assuming a fixed SN energy injection $W_{e, \rm S1}$ (Eq.~(\ref{Einj})) and $s=1.5$. Additionally, $E_{ke,\rm min}$ was fixed to ensure a smooth connection
between the PL and thermal distributions. For the given $W_{e,\rm S1}$ and $s$, the other parameters except for $n_i$ (and thus $n_{e,b}$; Eq.~(\ref{conserv})) are well constrained; $n_i$ exhibits flexibility within a broad range. This is because the emission, which is proportional to $n_i$, is counterbalanced by the effects of $n_{e,b}$ ($\sim n_i$) since emission (cooling) timescale is inversely proportional to $n_{e,b}$: $t_{\rm cool}\approx E_{ke}(0)^{1.5}/(1.16\times 10^{-5} \lambda_{ee} n_{e,b})$ (Eq.~(\ref{Ecoul})).

This cooling timescale has observational consequences. Under continuous shock acceleration over $\tau_{\rm age}$, the X-ray flux of S1 would initially rise for $t_{\rm cool}$ because the injection rate exceeds the cooling rate during the initial phase. The emission then remains stationary for the rest of its age, $\tau_{\rm age}-t_{\rm cool}$. We adjusted $n_{e,b}$ such that this stationary period is longer than the $\sim$15\,yr period of relatively constant observed X-ray flux (Table~\ref{ta:ta1}); a larger $n_{e,b}$ is also acceptable as it extends this period. Table~\ref{ta:ta3} shows a sample set of parameters.

It is important to note that alternative choices for the parameter values can also provide acceptable fits to the data due to parameter covariance (especially with $W_{e, \rm S1}$). Consequently, the specific values reported in Table~\ref{ta:ta3}
represent just one possible solution within a range of possibilities.
We present further discussions in Section~\ref{sec:sec4_3}.

\begin{table}[t]
\vspace{-0.0in}
\begin{center}
\caption{Parameters for the NTB model in Figure~\ref{fig:fig4}}
\label{ta:ta3}
\scriptsize{
\begin{tabular}{llcc} \hline\hline
Parameter        & Symbol      & Value     \\ \hline
SN energy ($10^{51}$\,erg)                  &   $E_{\rm SN}$         & 1\marka       \\
Solid angle fraction of S1                  &   $f_{\Omega, S1}$     & 0.013\marka   \\
Energy conversion efficiency                &   $\eta$               & 0.1\marka     \\
Injected energy ($10^{48}$\,erg)            &   $W_{e,\rm S1}$       & 1.3\marka     \\
Index of electron distribution              &   $s$                  & 1.5\marka     \\  
Minimum energy of electrons (eV)            &   $E_{ke,\rm min}$     & 3.5\marka     \\
Maximum energy of electrons (keV)           &   $E_{ke,\rm max}$     & 120    \\
Injected electron density ($\rm cm^{-3}$)   &   $n_e$                & 39.1\markb     \\
Background ion density ($\rm cm^{-3}$)      &   $n_i$                & 100    \\
Background electron density ($\rm cm^{-3}$) &   $n_{e,b}$            & 80.9\markb    \\  
Lifetime of S1 (yr)                         &   $\tau_{\rm age}$     & 55     \\ \hline
\end{tabular}}
\end{center}
\vspace{-0.5 mm}
\footnotesize{
\marka{Fixed.}\\
\markb{Initial values. These values evolve with time as injected electrons ($n_e$) cool and transition into the background population ($n_{e,b}$). See Section~\ref{sec:sec3_1} for details.}\\
}
\end{table}

\section{Discussion}
\label{sec:sec4}
The origin of the non-thermal X-ray emission from S1 has been controversial.
\citet{Nakamura2009} proposed that these X-rays share a common origin with the
radio shell emission of CTB~37B based on their similar spectral indices.
Alternatively, \citet{Blumer2019} suggested that a shock interaction with a nearby MC or
a PWN unassociated with CTB~37B could be responsible.
\citet{Gotthelf2019} attributed the emission to a background PWN
based on the high value of $N_{\rm H}$ that they inferred.

Leveraging improved photon statistics facilitated by
XMM-Newton PN and NuSTAR data, we obtained a refined measurement of the S1 spectrum.
Our analysis of the S1 X-ray spectrum revealed that 
three models (PL, {\tt srcut}, and BPL) provide adequate fits,
with the $F$-test favoring the BPL model. 
None of the three could be definitively excluded,
leaving an association between S1 and CTB~37B inconclusive.  
All three explanations have significant difficulties, as we describe below.

\subsection{Scenario 1: Emission from a PWN unrelated to CTB~37B}
\label{sec:sec4_1}
If the true emission spectrum of S1 is a PL, the inferred $N_{\rm H}$ of
$6.5\times 10^{22}\rm \ cm^{-2}$ is incompatible with those for the SNR and J1714
at a $>3\sigma$ level (Table~\ref{ta:ta1}).
This suggests that S1 is not physically associated with the SNR,
aligning with the interpretation of a background PWN proposed by \citet{Gotthelf2019}.
The properties of S1 ($L_X = 3.6 \times 10^{33} (d/10\ {\rm kpc})^2$ erg s$^{-1}$ and $\Gamma \sim 2$)
would be quite typical for a PWN \citep[see Figure 5 of][]{Kargaltsev2013}.
In this scenario, the non-detection by XMM-Newton or Chandra of a central point source,
potentially a middle-aged pulsar, is somewhat puzzling since existing empirical correlations
between luminosities of pulsars and their PWNe \citep[][]{Kargaltsev2008,Li2008} suggest
that pulsars should be as luminous as their PWNe. However, X-ray PWNe are
sometimes without detected pulsars. While most of the 91 X-ray PWNe tabulated in \citet{Kargaltsev2013} have a point source(s) within them, pulsations have not been detected in 15 of them, making the association between the PWN and the point source(s) unclear.
Moreover, the putative pulsar may be observationally faint due to strong absorption if its emission is spectrally soft.
Deeper X-ray observations with future instruments like AXIS and HEX-P \citep{Reynolds2023, Madsen2024}
could help resolve this issue.

\subsection{Scenario 2: Synchrotron Emission from the SNR Shock}
\label{sec:sec4_2}
Highly relativistic electrons, possibly accelerated by interaction between S1 and the SNR shock,
are capable of generating X-rays via synchrotron radiation.
In this case, one might expect a slowly cutting-off spectrum that can be described by the 
simple {\tt srcut} model, as is seen in other remnants \citep[e.g., Tycho;][]{Lopez2015}.
However, the rolloff photon energies reported in Table~\ref{ta:ta2} (5--10\,keV, depending on radio
properties) would be the highest ever observed for a SNR.  When electron acceleration is
limited by synchrotron losses, the characteristic rolloff photon energy is given by
\begin{equation}
\label{rolloff}
    h\nu_{\rm rolloff} \sim 2 \left( \frac{u_{\rm shock}}{1000\ {\rm km\ s}^{-1}} \right)^2
    \left(\eta_g R_J \right)^{-1} \ {\rm keV}
\end{equation}
\citep[e.g.,][]{Reynolds2008}.
Here $\eta_g \equiv \lambda_{\rm mfp}/r_g$ is the ``gyrofactor'', the electron mean free path in units of its gyroradius, and $R_J$ is a geometric factor reflecting potential variations in
acceleration rate as a function of the shock obliquity angle $\theta_{\rm Bn}$ between the
shock velocity and upstream magnetic field \citep[][]{Jokipii1987}.  \citet{Jokipii1987} shows that
as $\eta_g$ increases, acceleration can proceed much faster in perpendicular shocks ($\theta_{\rm Bn} \sim \pi/2$) than in parallel ones, though in an increasingly narrow range of $\theta_{\rm Bn}$
near $\pi/2$.  For large $\eta_g,$ $R_J$ varies as $\eta_g^{-2}$, so Equation~(\ref{rolloff}) shows
that higher rolloffs could be obtained invoking this effect with large values of $\eta_g$.
The relatively low shock velocity \citep[$\sim$900\,\kms;][]{Blumer2019} means that a very
large value of $\eta_g$ would be required.
IXPE \citep[][]{IXPE2022} observations of SNRs revealed small $\theta_{\rm Bn}$ values
in young SNRs \citep[][]{Slane2024}, potentially indicating a modest $\eta_g$ for S1.
However, examples of tangential (large $\theta_{Bn}$) magnetic fields also exist in older SNRs
\citep[e.g.,][]{Prokhorov2024}.
Polarization angle measurements for CTB~37B and S1 are
crucial to validate the feasibility of this scenario.

Nevertheless, the {\tt srcut} model implies a higher $N_{\rm H}$ than estimates for the SNR and J1714 at the 3$\sigma$ level, casting doubt on the association between S1 and CTB~37B.

\subsection{Scenario 3: NTB Emission from S1}
\label{sec:sec4_3}
The BPL model suggests a different scenario.
The consistency between $N_{\rm H}$ towards S1 and
that measured for the SNR and J1714, along with the absence of a point-like source within S1,
argues for a physical association with CTB~37B.
Furthermore, the hard spectral index below 6\,keV and the observed degree of the spectral break ($\Delta \Gamma \approx 1$)
favor a NTB interpretation over synchrotron radiation,
as synchrotron emission from PWNe is typically softer ($\Gamma\approx$2)
with smaller $\Delta \Gamma$ values \citep[e.g., $\le 0.5$;][]{Bamba2022}.
The NTB scenario aligns with the suggestion of \citet{Blumer2019} regarding an interaction
between the SNR shock and an MC.
While our NTB model provided a successful explanation for the X-ray measurements, this NTB scenario also has some difficulties in detail.

These difficulties stem primarily from the inefficiency of NTB radiation compared to Coulomb losses \citep[see][for details]{Petrosian2008} and the limited energy budget of the SN. Only a small fraction \citep[$\sim 10^{-5}$;][]{Petrosian2001} of electron energy contributes to NTB emission. Consequently, the available SN energy of $W_{e,\rm S1}\sim 10^{48}$\,erg can sustain the observed flux of $F_{2-10\rm \ keV} \sim 3\times 10^{-13}$\,\fluxcgs\ for only $\sim$100\,yrs, significantly shorter than the estimated SN age of 650--6200\,yr \citep[][]{Nakamura2009,Blumer2019}.
Our NTB model reflects this constraint by assuming that all available SN energy is dedicated to electron acceleration, resulting in $\tau_{\rm age} = 55$\,yr for S1 age. However, it remains unclear whether the interaction process accelerates only electrons (and not protons). If the SN energy is shared with protons, the age estimate decreases, potentially conflicting with the observed stability of the X-ray flux over $\approx$13.5\,yrs (Table~\ref{ta:ta1}).
Additionally, an asymmetric SN explosion (e.g., J1714's proximity to the western shell; Figure~\ref{fig:fig1})
could have injected less energy into S1, leading to a reduced $W_{e,\rm S1}$ and a smaller $\tau_{\rm age}$.
We note that a similar requirement of a young age arises from
the rather low value of break energy $E_{\rm brk}$, necessitating a small value
of $n_{e,b} \tau_{\rm age}$, as already remarked in Section~\ref{sec:sec3_2}.

It is important to acknowledge that our model considered only Coulomb cooling for
the evolution of the electron distribution.
In reality, additional processes, including turbulence and
magnetic field interactions \citep[e.g.,][]{Bykov2000}, likely play significant roles.
If these processes induce sufficiently rapid acceleration,
the non-thermal PL tail (responsible for NTB emission) may persist for extended periods
\citep[][]{Petrosian2008}, potentially mitigating the aforementioned challenges.
Further investigation into these complexities is warranted to develop a more
comprehensive model for the non-thermal emission from S1,
including predictions for Fe~K$\alpha$ flux from excitation by non-thermal electrons,
which XRISM \citep[][]{XRISM2020} can test.

\subsection{Possibility of TeV Emission from S1}
\label{sec:sec4_4}
Our analysis suggests three potential origins for the non-thermal X-ray emission from S1: an unassociated PWN (Scenario~1), the SNR shock accelerating relativistic, synchrotron-emitting electrons (Scenario~2), or the SNR shock accelerating suprathermal but non-relativistic electrons producing NTB (Scenario~3). 
Scenarios 1 and 2 predict IC emission at TeV energies or higher, although precise flux estimates are challenging.

In Scenario~1, the S1 size of $R\approx 1'$ ($\sim$4\,pc for an assumed distance of 13\,kpc scaled by $N_{\rm H}$)
may indicate a middle-aged PWN. Such a PWN could exhibit a TeV flux comparable
to its X-ray flux \citep[e.g.,][]{Park2023a}, which is an order of magnitude lower than
the measured gamma-ray flux of the SNR shell (peaking at $\sim$100\,GeV). IC emission from the electrons emitting synchrotron photons at $\sim$10\,keV would appear at $\ge$TeV energies, as observed in other middle-aged PWNe.
Similarly, the high rolloff energies of 5--10\,keV inferred from the {\tt srcut} model (Scenario~2) imply electron energies of $\ge 30$\,TeV for an assumed $B=100\mu$G within the SNR shock \citep[e.g.,][]{Zeng2017}. These electrons could upscatter ISRF (e.g., 30\,K blackbody) to $\ge$\,TeV.

In Scenario~3, S1 is unlikely to produce a significant gamma-ray flux on its own
due to the limited energy budget (i.e., lack of $>$TeV electrons).
However, the surrounding SNR shell could potentially contribute high-energy particles
to S1 via energetic proton diffusion. Similar scenarios involving particle escape
from SNR shells and interaction with nearby clouds have been proposed for sources
like SNR W28 \citep{Aharonian2008W28} and dark accelerators \citep{Gabici2009}.
While the small solid angle coverage of S1 ($f_{\Omega, S1}\approx 1.3$\%) limits
the number of protons reaching the cloud, the high gas density within S1
(estimated ion density of $100\rm\  cm^{-3}$; Table~\ref{ta:ta3})
could still potentially lead to detectable gamma-ray emission.
It is important to note that proton diffusion is energy-dependent \citep{Aharonian1996},
preferentially enriching S1 with higher-energy protons from the shell.

In summary, TeV emission from S1 is possible under any of the three scenarios for the hard X-ray emission.
In particular, S1 may manifest as a distinct ``high-energy'' TeV source distinguishable from the SNR shell itself. Future TeV observatories like the Cherenkov Telescope Array \citep[CTA;][]{CTA2011} may have the resolving power to reveal such a source.

\section{Summary}
\label{sec:sec5}
Despite our investigation, the origin of the non-thermal X-ray emission from S1 remains uncertain.
The X-ray data favor the BPL spectrum over the PL and {\tt srcut} ones, but the latter two cannot be definitively ruled out. These spectral models suggest three different scenarios (Sections~\ref{sec:sec4_1}--\ref{sec:sec4_3}) for the S1 emission, and we find that all three scenarios have significant problems. The NTB scenario (BPL model; Scenario~3) requires an unrealistically short source age, and its overall energetics are problematic.  The synchrotron explanation ({\tt srcut} model; Scenario~2) requires dramatically more rapid electron acceleration than has been documented in other sources, although in principle such acceleration cannot be ruled out.  The unrelated-PWN explanation (PL model; Scenario~1) probably has the fewest fatal
flaws, requiring only a somewhat unlikely spatial coincidence between a fairly conventional X-ray PWN and the CTB~37B shell.  The various scenarios are testable:
A detection of a central point source with deep Chandra observations
could lend credence to `unassociated PWN' scenario.
The NTB scenario predicts a decline in the X-ray flux over the next decades
due to Coulomb cooling (but see Section~\ref{sec:sec4_3}).
The keV electrons producing NTB should also excite atomic
lines, most prominently Fe K$\alpha$; detecting such emission from S1, e.g., with XRISM,
could strengthen the case for NTB while more stringent upper limits could weaken it.
Further X-ray observations are essential to distinguish among these scenarios.

\begin{acknowledgments}
This work used data from the NuSTAR mission, a project led by the California Institute of Technology,
managed by the Jet Propulsion Laboratory, and funded by NASA. We made use of the NuSTAR Data
Analysis Software (NuSTARDAS) jointly developed by the ASI Science Data Center (ASDC, Italy)
and the California Institute of Technology (USA).
JP acknowledges support from Basic Science Research Program through the National
Research Foundation of Korea (NRF) funded by the Ministry of Education (RS-2023-00274559).
This research was supported by the National Research Foundation of Korea (NRF)
grant funded by the Korean Government (MSIT) (NRF-2023R1A2C1002718). SSH’s research is primarily supported by the Natural Sciences and Engineering Research Council of Canada (NSERC)
through the Canada Research Chairs and the Discovery Grants programs.
\end{acknowledgments}

\bigskip
\vspace{5mm}
\facilities{XMM-Newton \citep[][]{jlac+01}, NuSTAR \citep[][]{hcc+13}}
\software{HEAsoft \citep[v6.31;][]{heasarc2014},
XMM-SAS \citep[20211130\_0941;][]{xmmsas17}, XSPEC \citep[v12.12;][]{a96}}


\bibliographystyle{apj}
\bibliography{ctb37b}
\end{document}